\documentclass[reprint,amssymb, amsmath, aps, superscriptaddress, showpacs, footinbib, prb]{revtex4-2}
\usepackage{graphicx,epstopdf} %Include figure files,superscriptaddress
\usepackage{amsmath}
\usepackage{color}
\usepackage{float}
\usepackage{soul}
\usepackage[hidelinks]{hyperref}

\newcommand{\be}{\begin{equation}}
\newcommand{\ee}{\end{equation}}
\newcommand{\bea}{\begin{eqnarray}}
\newcommand{\eea}{\end{eqnarray}}
\newcommand{\bse}{\begin{subequations}}
\newcommand{\ese}{\end{subequations}}

\begin{document}
\title{ Extrinsic to intrinsic mechanism crossover of anomalous Hall effect in the Ir-doped MnPtSn Heusler system}
\author{Sk Jamaluddin}
\affiliation{School of Physical Sciences, National Institute of Science Education and Research, An OCC of Homi Bhabha National Institute, Jatni-752050, India}
\author{Roumita Roy}
\affiliation{School of Physical Sciences, Indian Institute of Technology Goa, Ponda-403401, Goa, India}
\author{Amitabh Das}
\affiliation{ Solid State Physics Division, Bhabha Atomic Research Centre, Trombay Mumbai 400085, India}
\affiliation{Homi Bhabha National Institute, Anushaktinagar, Mumbai 400094, India}
\author{Sudipta Kanungo}
\email{sudipta@iitgoa.ac.in}
\affiliation{School of Physical Sciences, Indian Institute of Technology Goa, Ponda-403401, Goa, India}
\author{Ajaya K. Nayak}
\email{ajaya@niser.ac.in}
\affiliation{School of Physical Sciences, National Institute of Science Education and Research, An OCC of Homi Bhabha National Institute, Jatni-752050, India}

\date{\today}

\begin{abstract}

Recent findings of large anomalous Hall signal in nonferromagnetic and nonferrimagnetic materials suggest that the magnetization of the system is not a critical component for the realization of the anomalous Hall effect (AHE).  Here, we present a combined theoretical and experimental study demonstrating the evolution of different mechanisms of AHE in a cubic Heusler system MnPt$_{1-x}$Ir$_x$Sn. With the help of magnetization and neutron diffraction studies, we show that the substitution of nonmagnetic Ir in place of Pt significantly reduces the net magnetic moment from 4.17 $ \mu _B$/f.u. in MnPtSn to 2.78 $ \mu _B$/f.u. for MnPt$_{0.5}$Ir$_{0.5}$Sn. In contrast, the anomalous Hall resistivity is enhanced by nearly three times from 1.6~$ \mu \Omega $ cm in MnPtSn to about 5~$ \mu \Omega $ cm for MnPt$_{0.5}$Ir$_{0.5}$Sn. The power law analysis of the Hall resistivity data suggests that the extrinsic contribution of AHE that dominates in the case of the parent MnPtSn almost vanishes for MnPt$_{0.5}$Ir$_{0.5}$Sn, where the intrinsic mechanism plays the major role. The experimental results are well supported by our theoretical study, which shows a considerable enhancement of the spin-orbit coupling when Ir is introduced into the system. Our finding of a crossover of the anomalous Hall effect with chemical engineering is a major contribution toward the recent interest in controlling the band topology of topological materials, both in bulk and thin-film forms.

\end{abstract}

%\pacs{75.50.Gg, 75.50.Cc, 75.30.Gw, 75.70.Kw}
% PacS, the Physics and Astronomy Classification Scheme.
%Use showkeys class option if keyword display desired

\maketitle

%%%%%%%%%%%%%%%%%%%%%%%%%%%%%%%%%%%%%%%%%%%%%%%%%%%%%%%%%%%%%%%%%%%%%%%%%%%%%%%%%%%%%%%%%%%%%%%%%%%%%%%%%%%%%%%%%%%%%%%%%%%%%%%%%%%%%%%%%%%%%%%%%%
\section{INTRODUCTION}
In general, the anomalous Hall effect (AHE) in ferromagnetic (FM)/ferrimagnetic (FiM) materials scales with the magnetization of the samples. However, the recent observation of extremely large anomalous Hall conductivity in some of the moderate magnetization based FM materials \cite{Fe3Sn2,Co2MnGa Science,Co2MnAl} and nonmagnetic materials \cite{ZrTe5,Kv3Sb5} suggests that the spin-orbit coupling driven band topology is one of the primary microscopic mechanisms that governs the AHE. In the case of FM materials, the total Hall resistivity can be expressed as $\rho_{xy}= \rho^o_{xy}+\rho^A_{xy}= R_{0}H_{z}+R_{s}M_{z}$, where $R_0$ and $R_S$ are the ordinary and anomalous Hall coefficients, respectively. It is well established that mainly two different microscopic mechanisms, such as a scattering-dependent extrinsic contribution and scattering-independent intrinsic process, dictate the AHE in FM materials \cite{reviewNagaosa}. The extrinsic mechanism of AHE involves two types of scattering processes, the skew scattering \cite{7} and the side jump \cite{8}. Both of them are related to the scattering of charge carriers by the impurity sites and are the consequences of spin-orbit interaction. In contrast, the scattering-independent intrinsic mechanism proposed by Karplus and Luttinger arises due to the anomalous velocity of the charge carriers caused by the electronic band structure in the presence of spin-orbit coupling (SOC) \cite{KL}. Furthermore, the inclusion of the Berry phase to understand the intrinsic mechanism of AHE throws a deep insight regarding the band topology of the system \cite{9,Mn3Ge,Mn3Sn}. 

In real systems, different microscopic mechanisms of AHE have been understood in terms of a power law relation between anomalous Hall resistivity ($\rho^{A}_{xy}$) and longitudinal resistivity ($\rho_{xx}$). For the intrinsic contribution, $\rho^{A}_{xy}$ scales quadratically with the longitudinal resistivity $\rho_{xx}$, i.e., $\rho^A_{xy} \propto\rho^2_{xx}$ \cite{Lee,Pu}. In the case of extrinsic skew scattering, the $\rho^{A}_{xy}$ linearly scales with $\rho_{xx}$, i.e., $\rho^A_{xy} \propto \rho_{xx}$, while for the side jump the relation is quadratic in nature, i.e.,  $\rho^A_{xy}$ $\propto$ $\rho^2_{xx}$. These relations can be further generalized to a scaling law $\rho^A_{xy}\propto\rho^{\alpha}_{xx}$  to identify the dominant mechanism of AHE \cite{Co3Sn2S2_Fe}, where $\alpha$ = 2 and 1 correspond to intrinsic /  side jump and extrinsic skew scattering contributions, respectively \cite{Co3Sn2S2 Nat,Fe5Sn3, FeCr2Te4}. Since the intrinsic and side jump mechanisms exhibit a similar relationship with the longitudinal resistivity, it is difficult to differentiate them experimentally.
Recently, Tian et al \cite{Tian scalling} introduced a new scaling relation called the Tian-Ye-Jin ($TYJ$) model to separate the different contributions of AHE experimentally \cite{La,MnxGa/Pt,Mn1.5Ga,Co3Sn2S2_Ni PRL}. It has been shown that the side jump contribution is negligibly small in most of the metallic ferromagnetic systems \cite{Tian scalling,La}, except in some of the multilayer systems where interfacial scattering plays an important role \cite{Fe-Au,Co2MnGa NPJ}. The new scaling relation without considering the side jump mechanism can be expressed as
\begin{equation}
	\rho^A_{xy}= a\rho_{xx0} +b\rho^2_{xx}
\end{equation}
where $\rho_{xx0}$ is the residual resistivity. The first term corresponds to the extrinsic skew scattering and the second one deals with the intrinsic mechanism, where $ b $ is the intrinsic parameter.

%%%%%%%%%%%%%%%%%%%%%%%%%%%%%%%%%%%%%%%%%%%%%%%%%%%%%%%%%%%%%%%%%%%%%%%%%%%%%%%%%%%%%%%%%%%%%%%%%%%%%%%%%%%%%%%%%%%%%%%%%%%%%%%%%%%
\begin{figure*}[tb!]
	\includegraphics[angle=0,width=15.5 cm,clip]{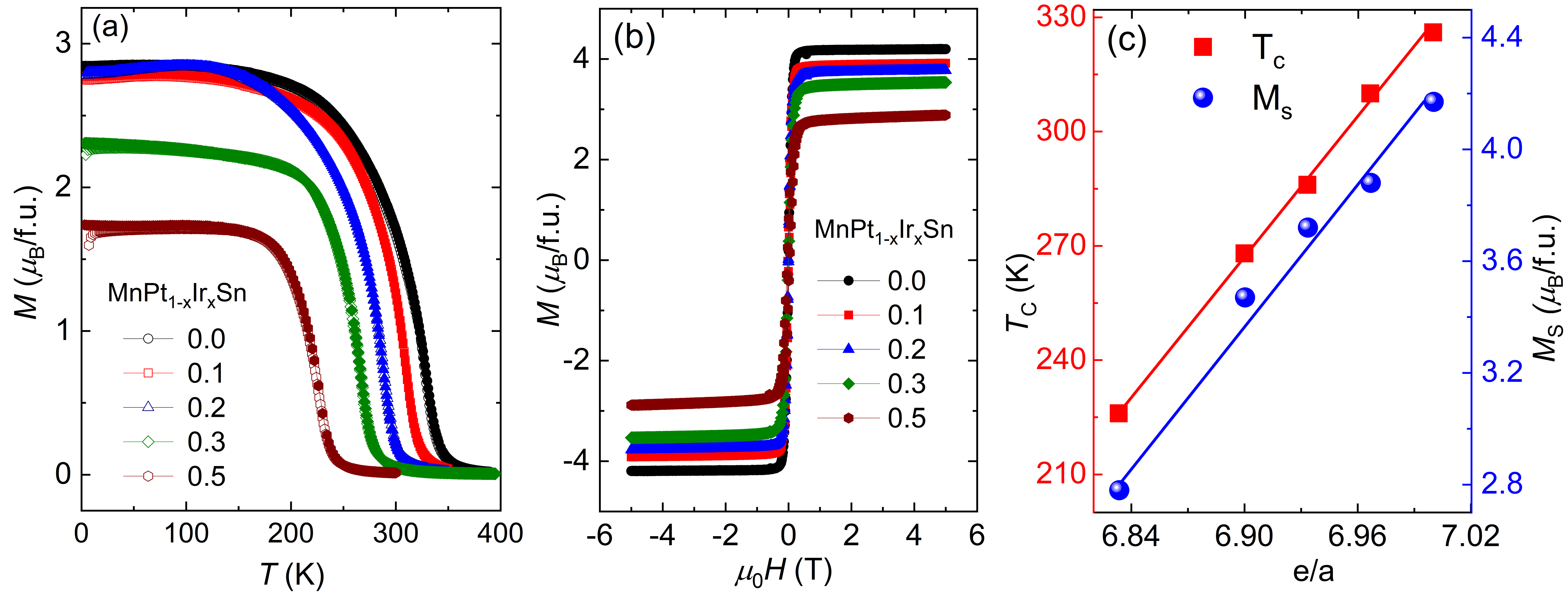}
	\caption{(a) Temperature-dependent zero-field-cooled (ZFC; open symbols) and field cooled (FC; closed symbols) magnetization [M(T)] curves for MnPt$_{1-x}$Ir$_x$Sn (x=0, 0.1, 0.2, 0.3, 0.5). (b) Field-dependent isothermal magnetization loops [$ M(H) $] measured  at 5 K for the MnPt$_{1-x}$Ir$_x$Sn samples. (c) Valence electron count (e/a) dependent Curie temperature and saturation magnetization for these samples.
		\label{MT}}
\end{figure*}
Although  AHE is a well studied phenomenon in several materials, a systematic manipulation of different mechanisms as discussed above is extremely important for its future implementation in real devices. In this regard, Heusler materials lay a great platform to willfully tune the material property by chemical doping \cite{Graf,Kubler,S4, CoTiFeSb}. In some of the recent studies, it is demonstrated that these materials can exhibit extremely large AHE due to the presence of Weyl nodes near the Fermi energy \cite{Weyl Node,HoPtBi}. Therefore, expanding the material base in the Heusler family with respect to the AHE and understanding its governing mechanism is an important step to move forward. In this report, we carry out a combined theoretical and experimental study to show a systematic change in the governing mechanism of AHE in cubic Heusler compounds  MnPt$_{1-x}$Ir$_x$Sn. First, we show that the magnetic moment of the system can be greatly modified by changing only the nonmagnetic element Pt and Ir. Then with the help of the scaling relation for the AHE and theoretical study, we present a detailed study on the changeover of the AHE mechanism from extrinsic to intrinsic by replacing Pt with Ir.

%%%%%%%%%%%%%%%%%%%%%%%%%%%%%%%%%%%%%%%%%%%%%%%%%%%%%%%%%%%%%%%%%%%%%%%%%%%%%%%%%%%%%%%%%%%%%%%%

\section{Methods}
Polycrystalline samples of MnPt$ _{1-x} $Ir$ _{x} $Sn (x=0, 0.1, 0.2, 0.3, 0.5) are prepared by arc-melting of ultrahighly pure constituent elements Mn, Pt, Ir, and Sn in argon atmosphere. For better chemical homogeneity, the ingots are melted four to five times by flipping the sides. As-prepared samples are annealed in an evacuated quartz tube at 850$^\circ$C for seven days, followed by quenching in ice water mixture. To check the phase purity, powder x-ray diffraction (XRD) measurement is performed on all the samples using a Rigaku SmartLab x-ray diffractometer with a Cu-K$_\alpha  $ source. The neutron diffraction patterns are recorded using the PD2 powder diffractometer ($\lambda$ = 1.2443 ${\AA}$) at the  Dhruva reactor, Bhabha Atomic Research Centre, Mumbai, India, at selected temperatures between 1.5 K and 300 K. The magnetic characterizations are carried out with the help of a Quantum Design MPMS3 (SQUID-VSM). The ac transport measurements are performed using a physical property measurement system (PPMS; Quantum Design).\\
The density-functional theory calculations are performed using the plane-wave basis and pseudopotential framework as implemented in the Vienna ${ab-initio}$ simulation package (VASP) \cite{Kresse1993,Kresse1996}. The exchange-correlation functional is employed following the
Perdew-Burke-Ernzerhof (PBE) prescription \cite{Perdew1996}. The experimentally determined lattice parameters are used in the calculations, while relaxing the atomic positions toward equilibrium until the Hellmann-Feynman force becomes less than 0.001 eV/$\AA$. In order to incorporate correlations beyond the scope of mean-field PBE, the Hubbard on-site $U$ is introduced by performing GGA+$U$ calculations \cite{Anisimov1993,Dudarev1998} with suitable values of U$_{eff}$ (U-J$_H$) of 5 eV at the Mn site and 2 eV at the Pt and Ir sites, respectively. The effect of SOC is introduced as a fully relativistic approach in the self-consistent calculations. The self-consistent electronic structure calculations are performed with a plane-wave cutoff of 500 eV, and an 8$\times8\times8$ k-mesh is used for the Brillouin zone integration.

\section{RESULTS AND DISCUSSION}

%%%%%%%%%%%%%%%%%%%%%%%%%%%%%%%%%%%%%%%%%%%%%%%%%%%%%%%%%%%%%%%%%%%%%%%%%%%%%%%%%%%%%%%%%%%%%%%%%%%%%%%%%%%%%%%%%%%%%%%%%%%%%%%%%%%%%%%%%%%%%%%%%

\begin{figure}[h]
	\centering
	\includegraphics[angle=0,width=7.3cm,clip]{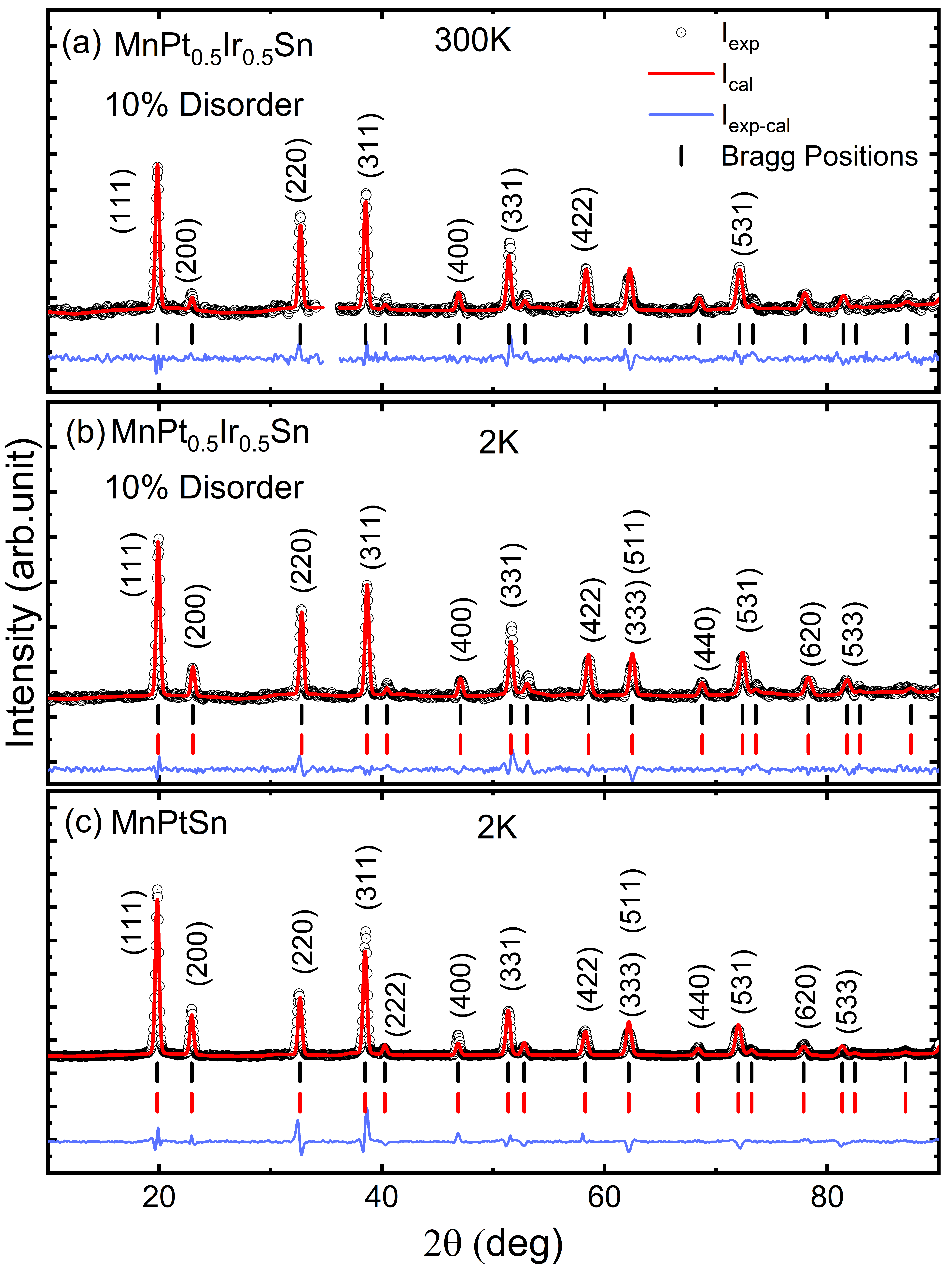}
	\caption{Rietveld refinement of powder neutron diffraction (ND) patterns. (a) Room-temperature Rietveld refinement of the powder ND pattern with 10\% atomic disorders between Mn/Ir site of MnPt$_{0.5}$Ir$_{0.5}$Sn. The region between 34 $^o$ and 36 $^o$ has been excluded due to contributions from the cryostat. (b) Low-temperature Rietveld refinement of powder ND pattern with 10 \% atomic disorder for  MnPt$_{0.5}$Ir$_{0.5}$Sn. (c) Low-temperature Rietveld refinement of the powder ND pattern for MnPtSn. 
		\label{Neu}}
\end{figure}
%%%%%%%%%%%%%%%%%%%%%%%%%%%%%%%%%%%%%%%%%%%%%%%%%%%%%%%%%%%%%%%%%%%%%%%%%%%%%%%%%%%%%%%%%%%%%%%%%%%%%%%%%%%%%%%%%%%%%%%%%%%%%%%%%%%%%%%%%%%%%%%%%%%%

\subsection{Structural and Magnetic Characterization}

The structural characterization for all the samples is carried out by room-temperature XRD measurements combined with Rietveld analysis using the FullProf suite (see Fig. S1 in the Supplemental Material) \cite{Suppli}. All the samples crystallize in a single cubic structure with space group $F\bar{4}3m$. A negligible change in the lattice parameter from $ a= $ 6.25 ${\AA}$ in MnPtIn to 6.23 ${\AA}$ is found with Ir doping for MnPt$_{0.5}$Ir$_{0.5}$Sn. The temperature-dependent magnetization, $ M (T) $, curves measured in zero-field-cooled (ZFC) and field-cooled (FC) modes for all the samples are shown in Fig. \ref{MT}(a). The magnetic ordering temperature, $ T_C $, for the parent  MnPtSn sample is found to be 326~K. The $ T_C $ decreases substantially with increasing Ir concentration and falls to 226~K for MnPt$_{0.5}$Ir$_{0.5}$Sn. The ZFC and FC  $ M (T) $ curves do not display any noticeable difference, signifying the absence of any spin-glass-like phase or considerable magnetic anisotropy in the system. The isothermal magnetization, $ M(H) $, loops measured at 5~K for all the samples  are depicted in Fig. \ref {MT}(b).  As can be seen, the saturation magnetization, $M_S$, significantly reduces from 4.17 $ \mu _B$/f.u. in the case of MnPtSn to about 2.78 $ \mu _B$/f.u. for MnPt$_{0.5}$Ir$_{0.5}$Sn. It is well known that the cubic $L2_1$ based Heusler materials show a linear variation of the magnetic moment with the valence electrons concentration \cite{Skaftouros,S5,cubic manganese}. In the present case, the substitution of Ir in place of Pt decreases the number of valence electron in the system. This is expected to greatly affect the electronic and magnetic properties of the system. Figure \ref {MT}(c)  shows the variation of total magnetic moment and $ T_C $ with the valence electron concentration (e/a). As can be seen, both $M_S$ and $ T_C $ exhibit a nearly linear dependency with the e/a ratio.

%%%%%%%%%%%%%%%%%%%%%%%%%%%%%%%%%%%%%%%%%%%%%%%%%%%%%%%%%%%%%%%%%%%%%%%%%%%%%%%%%%%%%%%%%%%%%%%%%%%%%%%%%%%%%%%%%%%%%%%%%%%%%%%%%%%%%%%%%%%%%%%%%%

 %%%%%%%%%%%%%%%%%%%%%%%%%%%%%%%%%%%%%%%%%%%%%%%%%%%%%%%%%%%%%%%%%%%%%%%%%%%%%%%%%%%%%%%%%%%%%%%%%%%%%%%%%%%%%%%%%%%%%%%%%%%%%%%%%%%%%%%%%%%%%%%%%%
 
To further understand the microscopic origin of the reduction in the total magnetic moment with Ir doping, we carry out a neutron diffraction (ND) study on our well-characterized polycrystalline samples MnPtSn and MnPt$_{0.5}$Ir$_{0.5}$Sn.  In the case of  MnPt$_{0.5}$Ir$_{0.5}$Sn, the ND measurements are performed at 2~K and 300~K. Since the $ T_C $ of MnPt$_{0.5}$Ir$_{0.5}$Sn is 226~K, we use the ND pattern at 300~K for this sample to extract the nuclear parameters. For this purpose, the Rietveld refinement using the FullProf suite \cite{Fullproof} is performed with the Wyckoff positions of Sn, Mn, and Pt/Ir as 4a (0, 0, 0), 4b ($\frac{1}{2}$, $\frac{1}{2}$, $\frac{1}{2}$), and 4c ($\frac{1}{4}$, $\frac{1}{4}$, $\frac{1}{4}$), respectively. However, we are unable to achieve a good fitting for the 300~K ND using the above mentioned Wyckoff positions. It is worth mentioning that the atomic disorder is frequently encountered in Heusler compounds with different transition metals of comparable ionic radii. Although Ir is substituted in place of Pt, the higher electronegativity difference between Ir and Sn in comparison to that of Pt-Sn may result in some of the Ir atoms sitting along with the Sn atoms \cite{Graf,Ni2MnSb}. To check this, we replace about 10\% of Mn atoms with Ir in the Mn Wyckoff position (4b) and put these Mn atoms in the Ir position (4c), as shown in Fig. \ref{Neu}(a). By doing so, we achieve a better fitting in comparison to that of without disorder [see Supplemental Material Fig. S2 (a)] \cite{Suppli}. In the case of 0\% disorder, the values of the $\chi^{2}$ and $R_{Bragg}$ factors are 6.40 and 13.74, respectively, whereas for 10\% atomic disorder the values of the $\chi^{2}$ and $R_{Bragg}$ factors are reduced to be 5.96 and 11.63, respectively. By further increasing the disorder, both the $\chi^{2}$ and $R_{Bragg}$ factors increase [see the Supplemental Material table S1]. Hence, it can be concluded from our Rietveld refinement that MnPt$_{0.5}$Ir$_{0.5}$Sn consists of about 10\% Mn-Ir atomic disorder. 

 \begin{figure*}[tb!]
 	\includegraphics[angle=0,width=15cm,clip]{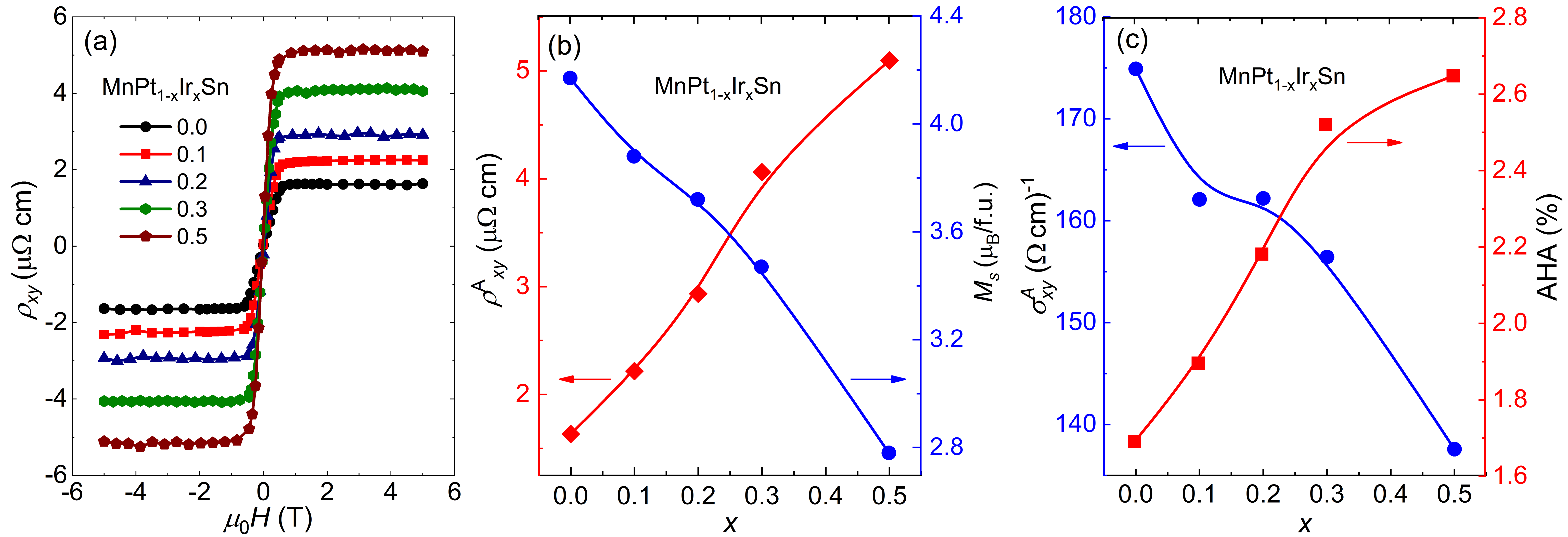}
 	\caption{(a) Field-dependent Hall resistivity [$\rho_{xy}(H)$] measured at 5 K for different samples. (b) Variation of anomalous Hall resistivity [$\rho_{xy}^A(H)$; red curve] and  saturation magnetization  ($ M_S $; blue curve) with iridium composition at 5 K for MnPt$_{1-x}$Ir$_x$Sn. (c) Compositional-dependent anomalous Hall conductivity ($\sigma^A_{xy}$) and anomalous Hall angle (AHA) at 5~K for  MnPt$_{1-x}$Ir$_x$Sn.
 		\label{fig2}}
 \end{figure*}
 
 Next, we concentrate on the low-temperature powder ND data to find out the magnetic ground state of our samples. In the case of  MnPt$_{0.5}$Ir$_{0.5}$Sn, the Mn atoms occupy both 4b and 4c sites due to the atomic disorder. First, we perform the refinement considering the magnetic moment at the 4b and 4c sites. The Mn moment at the 4b site comes to be 3.14 $\mu_B$, whereas the 4c site displays a very small moment. This may be due to the presence of a very small percentage of Mn atoms at the 4c site that makes it difficult to get the magnetic information. Then we carry out the refinement by giving the magnetic moment only at the 4b position.  This scenario does not affect the fitting parameter much and we obtain a magnetic moment of 3.17 $\mu_B$ for the Mn atoms. This results in a total moment of 2.85 $\mu_B/f.u.$  which is in good agreement with the experimental magnetic moment observed from the magnetization data. In the case of the MnPtSn sample, the $ T_C $ is about 326~K. We are unable to perform the ND experiment above room temperature; only 2~K  ND data are collected for this sample. Here, we consider the regular structure without disorder and the refined magnetic moment of Mn at the 4b site comes out to be 3.52 $\mu_B$. In general for Heusler compounds, the magnetic atoms sitting at the 4b and 4c sites align antiferromagnetically. Hence, the reduction in the magnetic moment for MnPt$_{0.5}$Ir$_{0.5}$Sn might be arising from the lower  Mn concentration at the 4b site as well as the antiferromagnetic alignment between Mn at the 4b and Mn at the 4c site.
 Hence, our ND study categorically establishes that the reduction of magnetic moment in the case of the Ir-doped samples arises mainly from the induced atomic disorder in the system.

%%%%%%%%%%%%%%%%%%%%%%%%%%%%%%%%%%%%%%%%%%%%%%%%%%%%%%%%%%%%%%%%%%%%%%%%%%%%%%%%%%%%%%%%%%%%%%%%%%%%%%%%%%%%%%%%%%%%%%%%%%%%%%%%%%%%%%%%%%%%%%%%%%

\subsection{Anomalous Hall Effect}

In order to explore the effect of Ir substitution on the electrical transport, we focus on a detailed Hall resistivity measurement on our well-characterized samples. The field-dependent Hall resistivity [$\rho_{xy}(H)$] measured at 5~K for different Ir-doped samples is shown in Fig. \ref {fig2}(a). As is found, $\rho_{xy}(H)$ exhibits a monotonic increment from about 1.6~$ \mu \Omega $ cm for MnPtSn to about 5~$ \mu \Omega $ cm for MnPt$_{0.5}$Ir$_{0.5}$Sn. It is worth noting here that the enhancement of the $\rho_{xy}(H)$ is observed despite a significant reduction in the magnetic moment from about 4.2~$ \mu_B $/f.u. for MnPtSn to 2.8~$ \mu_B $/f.u. for MnPt$_{0.5}$Ir$_{0.5}$Sn [Fig. \ref{fig2}(b)]. Although there is a three fold increase in the Hall resistivity, we observe a small decrease in the anomalous Hall conductivity $\sigma^A_{xy}$ with Ir doping [Fig. 3(c), left y-axis]. The decrease in the $\sigma^A_{xy}$ might be arising due to the increase in residual resistivity as a result of atomic disorder due to Ir doping. In contrast, the anomalous Hall angle (AHE) considerably increases with increasing the Ir concentration. This suggests that the Ir doping actually helps in the overall conversion of the longitudinal resistivity to Hall resistivity. 
\begin{figure}[tb!]
	\includegraphics[angle=0,width=8.5cm,clip]{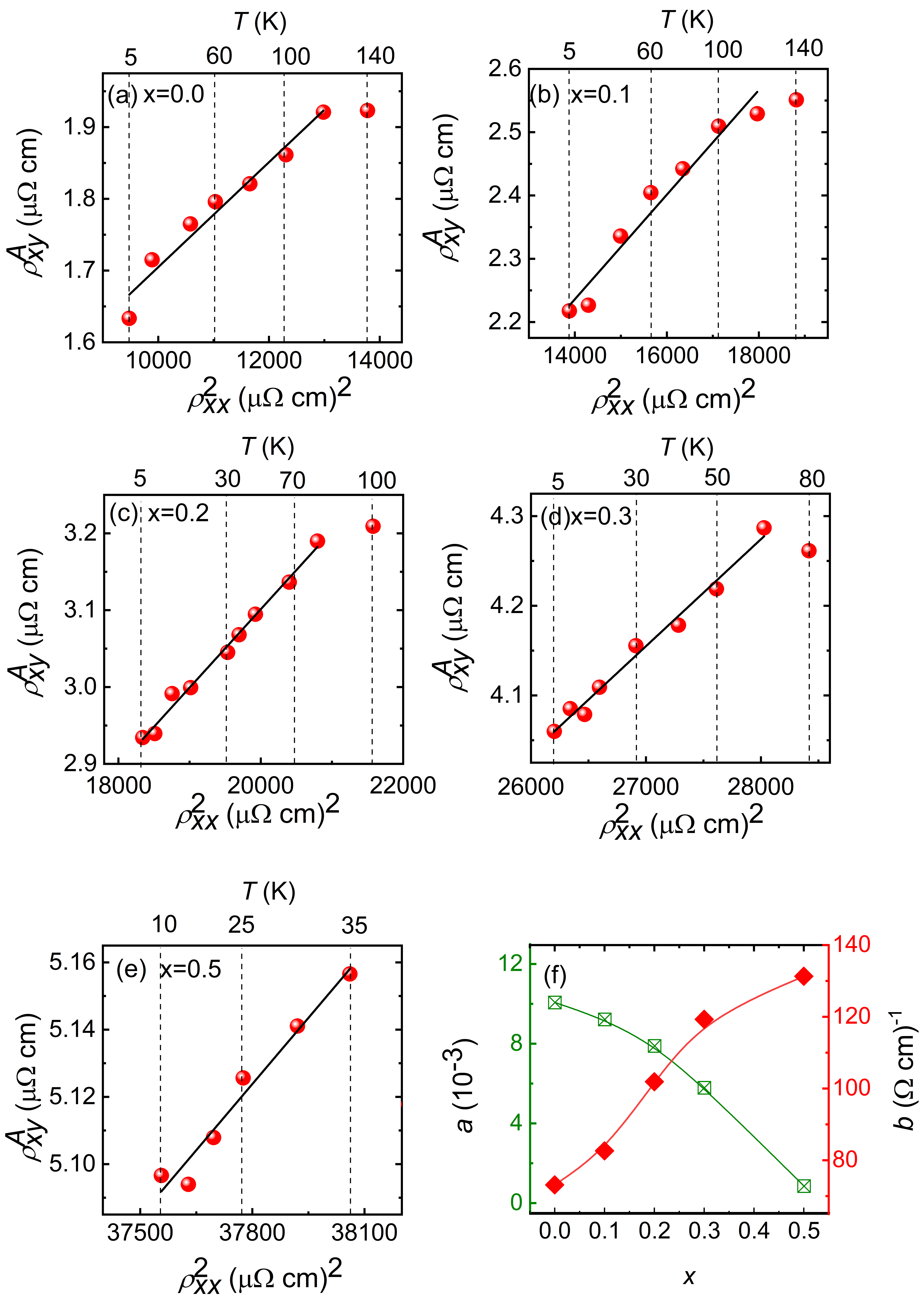}
	\begin{center}
		\caption{\label{fig5} (a-e) Anomalous Hall resistivity ($\rho^A_{xy}$) as a function of square of longitudinal resistivity ($\rho^2_{xx}$) for  MnPt$_{1-x}$Ir$_{x}$Sn (x= 0, 0.1, 0.2, 0.3, 0.5). Solid lines are the linear fitting to the equation $\rho^A_{xy}=a\rho_{xx0}+b\rho^2_{xx}$. Top axis indicates the corresponding measurement temperatures. (f) The extrinsic parameter ($ a $; open squares) and intrinsic parameter ($ b $; closed squares) as a function of iridium concentration $ x $. The lines are guides to the eye.}	
	\end{center}
\end{figure}
%%%%%%%%%%%%%%%%%%%%%%%%%%%%%%%%%%%%%%%%%%%%%%%%%%%%%%%%%%%%%%%%%%%%%%%%%%%%%%%%%%%%%%%%%%%%%%%%%%%%%%%%%%%%%%%%%%%%%%%%%%%%

%%%%%%%%%%%%%%%%%%%%%%%%%%%%%%%%%%%%%%%%%%%%%%%%%%%%%%%%%%%%%%%%%%%%%%%%%%%%%%%%%%%%%%%%%%%%%%%%%%%%%%%%%%%%%%%%%%%%%%%%%%%%

To understand the mechanism that governs the observed change in the AHE with Ir doping, we carry out a thorough analysis of the  $\rho^A_{xy}$ data using $ TYJ $ scaling relations \cite{Tian scalling} and the power law relation.
In order to map out the magnitude of different contributions of AHE in the Ir-doped samples, we employ the scaling relation given in Eq. (1). For this purpose, we plot $\rho^A_{xy}$ vs $\rho^2_{xx}$ for different samples, as depicted in Figs. \ref{fig5}(a)-\ref{fig5}(e). The $\rho_{xx}$ data are obtained from the temperature-dependent longitudinal resistivity measurements and $\rho^A_{xy}$ points are taken from the field-dependent Hall resistivity measurements at different temperatures as shown in the Supplemental Materials \cite{Suppli}.
The fitting parameters obtained from the linear fit [as shown by black solid lines in Figs. \ref{fig5}(a)-\ref{fig5}(e)] of the experimental data can be used to extract the information about different contributions of AHE.  The slope of the linear fitting gives the intrinsic parameter $ b $, whereas the extrinsic parameter $ a  $ can be calculated from the intercept. Figure \ref{fig5}(f) shows the variation of  $ a $ and $ b $ with the iridium concentration. The extrinsic parameter $ a $ decreases and the intrinsic parameter $ b $ increases with increasing iridium concentration. This illustrates that  the Ir doping enhances the intrinsic contribution to the AHE in the present system.  

It has been shown that the side jump contribution ($\sigma_{xx}^{SJ} $) is related to $\frac{e^2}{ha}(\frac{\epsilon_{so}}{E_{F}})$ \cite{extrinsic-to-intrinsic,S6}, where $\epsilon_{so}$ is the spin-orbit interaction energy, $E_F$ is the Fermi energy, and $ a $ is the lattice constant. Using this expression, we have calculated the approximate side jump contribution using $\frac{\epsilon_{so}}{E_{F}} \sim 0.01$ and the experimental lattice constant $ a $. We find that the side jump contribution for MnPtSn is almost negligible when compared with the extrinsic and intrinsic parameters \cite{Suppli}. For MnPt$_{0.5}$Ir$_{0.5}$Sn, the most dominant contribution arises from the intrinsic part, which is much larger than the side jump effect. However, the magnitudes of skew scattering and side jump contributions for this sample are comparable as both components are very small. Hence the extrinsic parameter `a' mainly represents the skew scattering. We summarize the values of $ a $, $ b $, anomalous Hall resistivity arising from the extrinsic contribution  $\rho^{ext}_{xy}$, the intrinsic anomalous Hall resistivity  $\rho^{int}_{xy}$, and the total $\rho^{tot}_{xy}$ for all the compounds in table 1.
%%%%%%%%%%%%%%%%%%%%%%%%%%%%%%%%%%%%%%%%%%%%%%%%%%%%%%%%%%%%%%%%%%%%%%%%%%%%%%%%%%%%%%%%%%%%%%%%%%%%%%%%%%%%%%%%%%%%%%%%%%%%%

\begin{figure}[tb!]
	\includegraphics[angle=0,width=8.5 cm,clip]{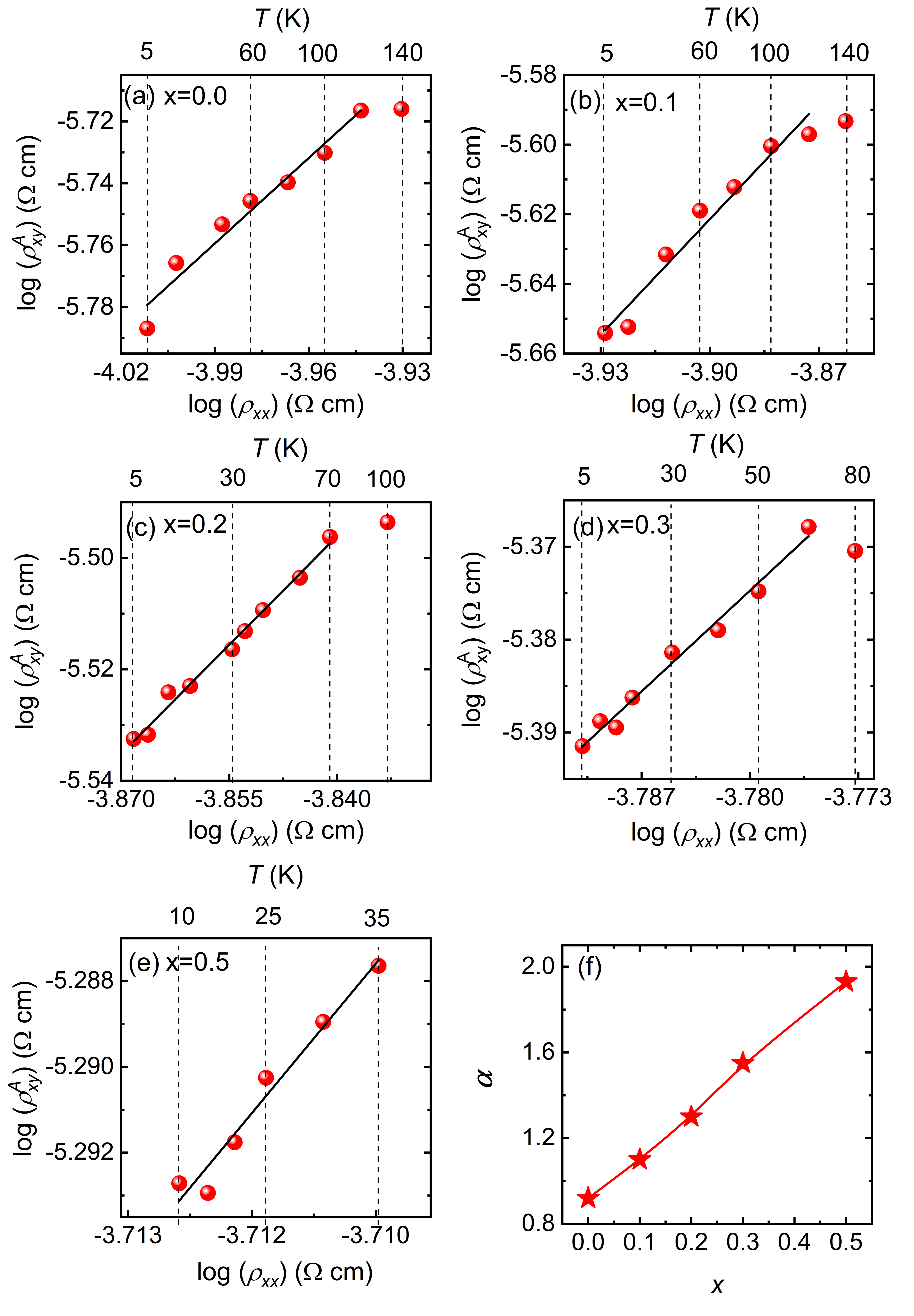}
	\caption{\label{log} Power law representation. (a-e) Plot of log ($\rho^A_{xy}$) vs log ($\rho_{xx}$) for  MnPt$_{1-x}$Ir$_{x}$Sn (x= 0, 0.1, 0.2, 0.3, 0.5). Black solid lines are the fit using the relation ($\rho^A_{xy}\propto\rho^{\alpha}_{xx}$). Top axis indicates the corresponding measurement temperatures. (f) Scaling factor $\alpha$ as a function of iridium concentration $ x $.}
\label{log}	
\end{figure}

%%%%%%%%%%%%%%%%%%%%%%%%%%%%%%%%%%%%%%%%%%%%%%%%%%%%%%%%%%%%%%%%%%%%%%%%%%%%%%%%%%%%%%%%%%%%%%%%%%%%%%%%%%%%%%%%%%%%%%%%%%%%
\begin{table}[tb!]
	\begin{center}
		\caption{Values of extrinsic and intrinsic parameters and different contribution of AHE.}
		------------------------------------------------------------------------------
		\label{tab:table1}
		\begin{tabular}{p{.25in}p{.5in}p{.7in}p{.47in}p{.65in}p{.45in}}\\

			\textbf{x} & \textbf{a } & \textbf{b}& \textbf{$\rho^{ext}_{xy}$ } & \textbf{$\rho^{int}_{xy}$} & \textbf{$\rho^{tot}_{xy}$}\\
			$  $ & $\times 10^{-3}$ & $ (\Omega$ $ cm)^{-1}$ &&$ (\mu\Omega$ $ cm)$   & \\
			\hline
			0.0      &     10.07        &     73.14		 &	0.97		&	0.68	& 1.65 \\ 
			0.1    &     9.22         &     82.60	     &	1.07		&	1.13	& 2.20\\ 
			0.2    &     7.89         &     101.9	     &	1.06		&	1.84	& 2.90\\ 		
			0.3    &     5.79         &     119.3        &	0.93		&	3.09	& 4.05\\ 
			0.5    &     0.856        &     131.1	     &	0.16		&	4.86    & 5.02\\ 
			\hline
		\end{tabular}
	\end{center}
\end{table}

 For the power law analysis, we plot log $\rho^A_{xy}$ vs log $\rho_{xx}$  as shown in Figs. \ref{log}(a)-\ref{log}(e). The solid lines represent the linear fitting to the data and the slope gives the scaling factor $\alpha$. For each sample a good fit to the data can be obtained up to a certain temperature below which the magnetization is almost constant.  As is found, $\alpha$ increases linearly with Ir doping from 0.96 for MnPtSn to 1.93 in MnPt$_{0.5}$Ir$_{0.5}$Sn [Fig. \ref{log}(f)]. The scaling factor $\alpha$ close to 1 suggests that the extrinsic contribution is the  dominant mechanism that governs the AHE in MnPtSn, whereas $\alpha$ of nearly 2 in the case of MnPt$_{0.5}$Ir$_{0.5}$Sn points toward its intrinsic nature. The change of $\alpha$ from 0.96 to 1.93 implies a changeover in the mechanism of AHE from extrinsic to intrinsic with Ir doping.

 \begin{figure*}[tb!]
 	\includegraphics[angle=0,width=15cm,clip]{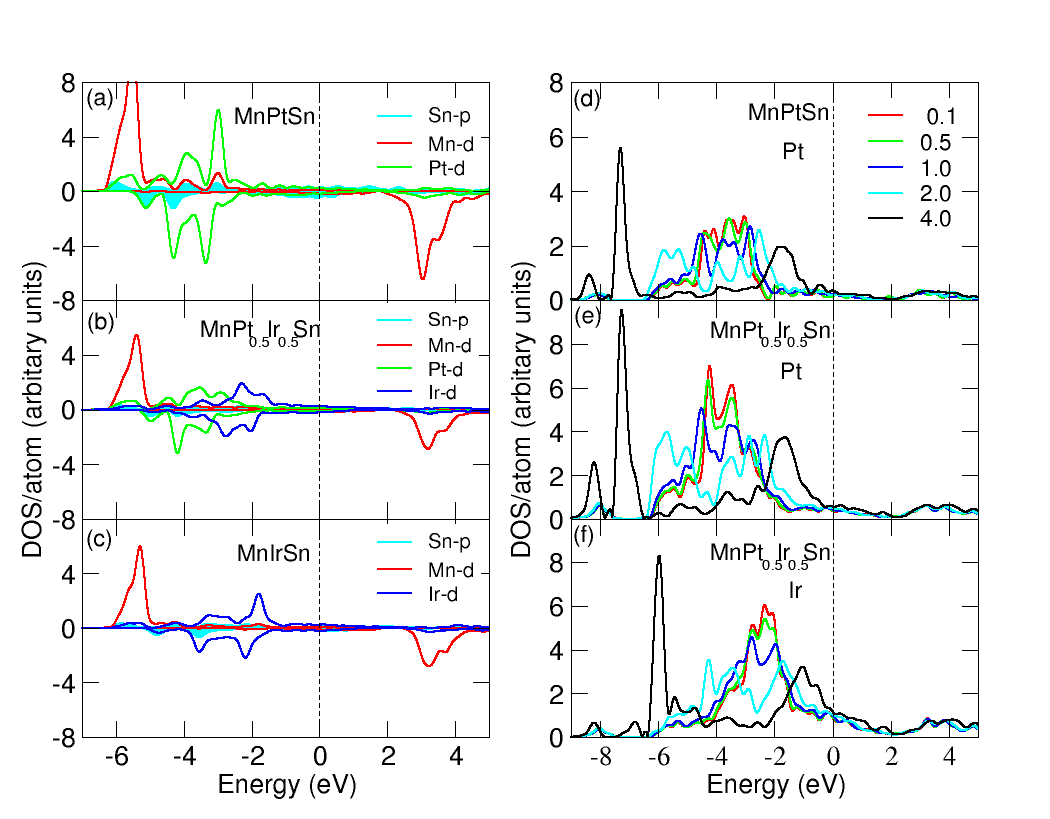}
 	\caption{ Projected density of states (DOS) with U$^{Mn}$ = 5eV and U$^{Pt/Ir}$ = 2eV for (a) MnPtSn, (b)MnPt$_{0.5}$Ir$_{0.5}$Sn, and (c) MnIrSn. The Mn-$\textit{d}$, Pt-$\textit{d}$, Ir-$\textit{d}$, and Sn-$\textit{p}$ states are represented by red, green, blue, and cyan, respectively. Projected density of states with the variation of spin-orbit coupling (SOC) strength. (d), (e), and (f) represent the variation of DOS for Pt-$\textit{d}$ in MnPtSn, Pt-$\textit{d}$ in MnPt$_{0.5}$Ir$_{0.5}$Sn, and Ir-$\textit{d}$ in MnPt$_{0.5}$Ir$_{0.5}$Sn, respectively. The color scheme for various values of $\alpha$ is mentioned in the right panel. The Fermi level in the energy scale is set at zero.} 
 		\label{MnPtSn_dos}
 \end{figure*}
 
 %%%%%%%%%%%%%%%%%%%%%%%%%%%%%%%%%%%%%%%%%%%%%%%%%%%%%%%%%%%%%%%%%%%%%%%%%%%%%%%%%%%%%%%%%%%%%%%%%%%%%%%%%%%%%%%%%%%%%%%%%%%

\subsection{Electronic Band Structure and Effect of the Spin-Orbit Coupling (SOC)}

%%%%%%%%%%%%%%%%%%%%%%%%%%%%%%%%%%%%%%%%%%%%%%%%%%%%%%%%%%%%%%%%%%%%%%%%%%%%%%%%%%%%%%%%%%%%%%%%%%%%%%%%%%%%%%%%%%%%%%%%%%%

The electronic structure calculations are carried out for three compounds in the MnPt$_{1-x}$Ir$_{x}$Sn series with x=0.0, 0.5 and 1.0. Although MnIrSn could not be synthesized experimentally, the crystal structure is obtained by the structural optimization starting from the MnPtSn and included in the study for the sake of complete understanding of the role played by each component. Figure \ref{MnPtSn_dos} shows the calculated GGA+$U$ comparative density of states (DOS) for the above-mentioned compounds. The DOS shows that the Mn-$\textit{3d}$ states are fully filled in the majority spin channel and completely empty in the minority spin channel, with a very small contribution at the Fermi energy. The major contribution at the Fermi level arises from the Pt-$5d$ and the Sn-$5p$ states as shown in Fig. \ref{MnPtSn_dos}(a).

The DOS clearly shows the metallic character of the electronic structure, where the Pt-$5d$ states contribute at the Fermi energy for MnPtSn. In the case of x=0.5, the hybridized Ir-$5d$ and Pt-$5d$ states mostly appear at the Fermi energy. The calculated magnetic moment at the Mn site comes out to be about 4.52 $\mu_B$, while the induced moments at the Sn and Pt sites appear to be -0.12 $\mu_B$ and negligibly small, respectively. This results in a total moment of 4.53 $\mu_B$/f.u. After 50\% Ir doping at the Pt site the DOS changes significantly as shown in Fig. \ref{MnPtSn_dos}(b). The Mn-$3d$ states remain almost unchanged, while the major renormalization happened at the Pt states due to incorporation of the Ir doping. Doped Ir-$5d$ states are pushed toward the Fermi level in MnPt$_{0.5}$Ir$_{0.5}$Sn. The magnetic moment at the Mn site is slightly reduced to 4.50 $\mu_B$, while the induced moment at the Ir site is around -0.05 $\mu_B$. The difference in the theoretical magnetic moment compared to the experimental moment obtained for the Ir doped sample might be arising due to the fact that the present theoretical calculations are performed without considering any disorder. For MnIrSn, the Mn-$3d$ states remain unchanged except slightly narrowing down the band width, and the Ir-$5d$ states are further pushed toward the Fermi level as shown in Fig. \ref{MnPtSn_dos}(c). Moreover, we calculate the magnetic transition temperatures in a mean-field method for the MnPtSn and MnPt$_{0.5}$Ir$_{0.5}$Sn. The calculated ratio of the transition temperature ($\frac{T^{MPS}_{c}}{T^{MPIS}_{c}}$) is around 1.52, which is in good agreement with the experimental value of 1.44 (see the Supplemental Material sec VII) \cite{Suppli}.

%%%%%%%%%%%%%%%%%%%%%%%%%%%%%%%%%%%%%%%%%%%%%%%%%%%%%%%%%%%%%%%%%%%%%%%%%%%%%%%%%%%%%%%%%%%%%%%%%%%%%%%%%%%%%%%%%%%%%%%%%%%%%%%%%%%%%%%%%%%%%%% 
\begin{figure*}[tb!]
	\includegraphics[angle=0,width=16cm,clip]{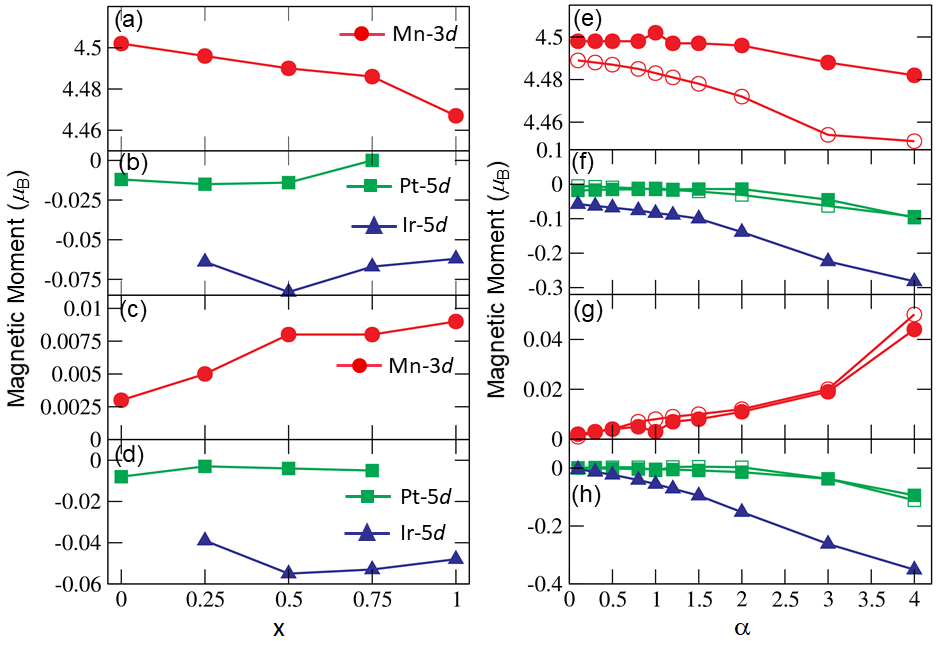}
	\caption{Variation of magnetic moment with respect to (a)-(d) iridium doping concentration (x)  and  (e)-(h) scale factor of SOC strength $\alpha$. (a),(b),(e),(f) represent the variation of spin magnetic moment and (c),(d),(g),(h) represent the variation of orbital magnetic moment. Mn-$\textit{d}$, Pt-$\textit{d}$, Ir-$\textit{d}$ states are shown by circle, square, and triangle symbols, respectively. In (e)-(h) closed symbols represent MnPtSn and open symbols MnPt$_{0.5}$Ir$_{0.5}$Sn samples.}	
	\label{Mag}
\end{figure*}

%%%%%%%%%%%%%%%%%%%%%%%%%%%%%%%%%%%%%%%%%%%%%%%%%%%%%%%%%%%%%%%%%%%%%%%%%

In addition to the above-discussed effects, Ir doping may also modify the effective SOC strength in the material. To capture the effect of SOC-induced modifications in the system, we have done electronic structure calculations for MnPtSn and MnPt$_{0.5}$Ir$_{0.5}$Sn by varying the SOC strength manually. The calculated GGA+$U$+SOC electronic DOSs with varying SOC strength ($\alpha$) for Pt/Ir states are shown in Figs. \ref{MnPtSn_dos}(d)-\ref{MnPtSn_dos}(f). As expected, the Mn-$3d$ state does not exhibit any substantial variation with the SOC strength \cite{Suppli}; however, the Pt/Ir-$5d$ states change significantly and move toward the Fermi energy with increasing SOC strength as shown in Figs. \ref{MnPtSn_dos}(d)-\ref{MnPtSn_dos}(f). As discussed above, a similar kind of modification of the DOS is also observed with Ir doping in Figs. \ref{MnPtSn_dos}(a)-\ref{MnPtSn_dos}(c). From these DOS calculations, it is very clear that the introduction of  Ir into the system substantially changes the effective strength of the SOC as well as the Pt/Ir states.

To understand the variation of the magnetic moment with the Ir doping, we also perform the calculations for intermediate concentrations (x= 0.25, 0.5, 0.75). The variation of spin and orbital magnetic moments with Ir concentration is shown in Figs. \ref{Mag}(a)-\ref{Mag}(d). It is evident that the spin magnetic moment at the Mn sites gradually decrease with increasing Ir concentration, whereas the Pt/Ir spin moments, although very small, remain almost constant. The point to be noted here is that the sign of the induced moments at the Ir site is opposite to that of the Mn site. As expected, the Mn atom exhibits a very small orbital moment in comparison to its spin magnetic moment and the orbital moment at the Pt/Ir site. Interestingly, the orbital magnetic moment at the Ir site is larger than that of the Pt site with opposite sign. However, the spin and orbital magnetic moments at the Ir sites follow almost the same trend with the Ir doping , as expected due to the modification of the effective strength of the SOC in the system.

%%%%%%%%%%%%%%%%%%%%%%%%%%%%%%%%%%%%%%%%%%%%%%%%%%%%%%%%%%%%%%%%%%%%%%%%%%%%%%%%%%%%%%%%%%%%%%%%%%%%%%%%%%%%%%%%%%%%%%%%%%%%%%%%%%%%%%%%
%%%%%%%%%%%%%%%%%%%%%%%%%%%%%%%%%%%%%%%%%%%%%%%%%%%%%%%%%%%%%%%%%%%%%%%%%%%%%%%%%%%%%%%%%%%%%%%%%%%%%%%%%%%%%%%%%%%%%%%%%%%%%%%%%%%%%%%%
\begin{figure*}[tb!]
	\includegraphics[angle=0,width=16cm,clip]{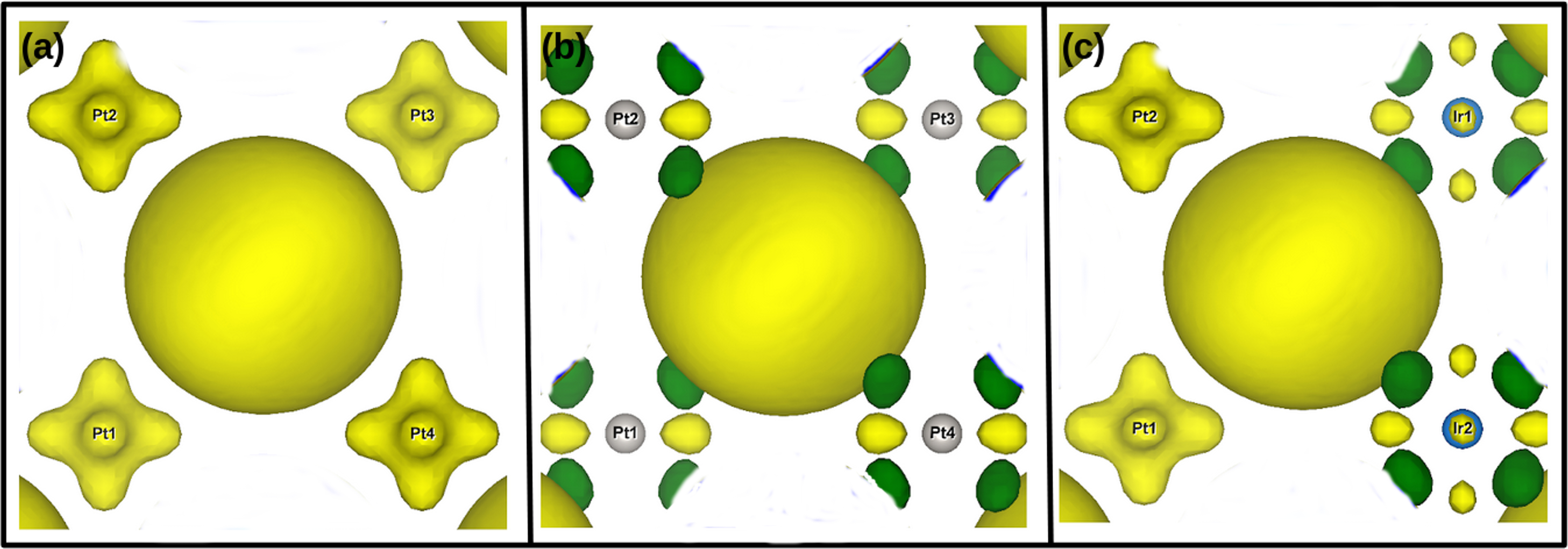}
	\caption{Magnetization density for the MnPtSn and 50\% Ir doped material for different SOC strength $\alpha$. (a) and (b) are the magnetization density of MnPtSn for $\alpha$=1.0 and 4.0 respectively, whereas (c) represents the same for MnPt$_{0.5}$Ir$_{0.5}$Sn with $\alpha$=1.0. Yellow and green represent the isosurfaces with + and - signs respectively.
		\label{mag_den}}
\end{figure*}

%%%%%%%%%%%%%%%%%%%%%%%%%%%%%%%%%%%%%%%%%%%%%%%%%%%%%%%%%%%%%%%%%%%%%%%%%%%%%%%%%%%%%%%%%%%%%%%%%%%%%%%%%%%%%%%%%%%%%%%%%%%%%%%%%%%%%%%%%%%%

From our experimental and theoretical studies, it is very clear that Ir doping plays a crucial role in deciding the magnetic state of the material. Naively, the physical consequence of the Ir doping at the place of Pt can be mapped as reducing one electron per atom in the material. As we can see from the DOS calculation, Ir doping changes the SOC which influences the resultant magnetic moment of the system. Figures \ref{Mag}(e)-\ref{Mag}(h) show the variation of spin and orbital magnetic moment as a function of the scale factor ($\alpha$) of the SOC strength. As the $\alpha$ increases, the spin magnetic moment at the Mn sites [Fig. \ref{Mag}(e)] decreases, whereas the orbital magnetic moment [Fig. \ref{Mag}(g)] increases. This trend can be understood from the transfer of spin contribution to the orbital contribution of magnetic moment with increasing SOC. In contrast, the variation of the spin and orbital moments as a function of $\alpha$ are very different for Pt and Ir. There is a marginal change in the spin and orbital magnetic moment for Pt as a function of $\alpha$, whereas the variation is quite substantial for Ir. The effect of SOC is more prominent in the case of Ir than Pt in both the MnPtSn and 50$\%$ Ir doped cases. Interestingly, the variations of magnetic moment at the Mn site as a function of the Ir doping and SOC strength ($\alpha$) are very similar, which suggests that the Ir doping at the Pt site plays a crucial role in dictating the effective SOC of the materials, apart from injecting an electron in the system.

%%%%%%%%%%%%%%%%%%%%%%%%%%%%%%%%%%%%%%%%%%%%%%%%%%%%%%%%%%%%%%%%%%%%%%%%%%%%%%%%%%%%%%%%%%%%%%%%%%%%%%%%%%%%%%%%%%%%%%%%%%%%%%%%%%%%%%%%%%%%%

\subsection{Magnetization Density}
The effect of the different roles played by the Pt and Ir atoms in terms of SOC strength can be conclusively visualized by the magnetization density plot shown in Fig.
\ref{mag_den}. Figures \ref{mag_den}(a) and \ref{mag_den}(b) show the magnetization density for MnPtSn with $\alpha$=1.0 and 4.0, respectively, whereas Fig.
\ref{mag_den}(c) represents the magnetization density for MnPt$_{0.5}$Ir$_{0.5}$Sn with $\alpha$=1.0. Since the Mn magnetic moment is much larger, the spin isosurface sitting at the Mn site is very large compared to the Pt/Ir sites. However, the small isosurface of the Pt or Ir carries very important information about the effective strength of the SOC. The Pt isosurface in Figs. \ref{mag_den}(a) and \ref{mag_den}(b), i.e. for $\alpha$=1.0 and 4.0, are very different, which is quite expected as the SOC strength modifies the shape of the magnetization density.
 A very interesting fact is that the shape of the Ir isosurface in Fig. \ref{mag_den}(c) (i.e, for MnPt$_{0.5}$Ir$_{0.5}$Sn with $\alpha$=1.0) is similar to that of the Pt isosurface in Fig. \ref{mag_den}(b) [i.e, for MnPtSn with $\alpha$=4.0]. 
This observation suggests that the effective strength of the SOC gets modified due to the Ir doping in the MnPtSn. Qualitatively, it can be said that the effective strength of the SOC in the case of the 50$\%$ Ir doped sample is greater than the parent MnPtSn. Hence, it is expected that the change in the SOC with Ir doping may significantly affect the electron transport in the system. The above discussed theoretical result nicely corresponds with the experimental findings. It is well established that the intrinsic contribution of the AHE is directly proportional to the relevant local Berry curvature of the material. To realize a non-vanishing Berry curvature by the breaking of the symmetry of the system, the SOC in the material plays a very crucial role \cite{XiaoRMP}. From the electronic structure calculations, we show that Ir doping enhances the effective strength of SOC; as a result the intrinsic contribution to the AHE should also increase. 

%%%%%%%%%%%%%%%%%%%%%%%%%%%%%%%%%%%%%%%%%%%%%%%%%%%%%%%%%%%%%%%%%%%%%%%%%%%%%%%%%%%%%%%%%%%%%%%%%%%%%%%%%%%%%%%%%%%%%%%%%%%%%%%%%%%%%%%%%%%%%

  The theoretical findings corroborate our experimental observation of enhanced AHE in case of the Ir-doped samples. It may be noted that the Ir-doped samples possess some amount of disorder as well as higher residual resistivity compared to the parent MnPtSn compound.  Although the skew scattering contribution is sensitive to the impurity, it is larger in a clean regime [low resistivity, $\sigma_{xx}>10^6$ $  (\Omega$ $  cm)^{-1}$] and decays in a bad metal regime [high resistivity, $ \sigma_{xx} <10^4$ $ (\Omega$ $cm)^{-1})$]. This suggests that as the conductivity $(\sigma_{xx})$ decreases, the skew scattering contribution starts diminishing and the intrinsic mechanism begins to dominate \cite{reviewNagaosa, Fe extrinsic}. In the present case all the samples fall into moderate to bad metal regimes. Hence, our experimental and theoretical findings suggest that one can tune different contributions of AHE with changing the overall SOC strength in the system. In this regard, AHE driven by different underlying mechanisms is reported recently in several materials. In some cases, a large intrinsic anomalous Hall signal has been observed due to the presence of  Weyl nodes \cite{Co2TiSn,GdPtBi}. Additionally, nontrivial scalar spin chirality can also induce a large extrinsic anomalous Hall effect as recently found in MnGe \cite{MnGe}. Besides this, there are reports of modification of intrinsic and extrinsic AHE by tuning the Fermi level with the help of chemical doping \cite{La,PrAlGe, FePt}. Although SOC is the primary mechanism responsible for the AHE, it is important to understand the correlation among other mechanisms contributing its cause.  In the present study, we demonstrate the manipulation of different contributions of AHE by introducing chemical doping.

%%%%%%%%%%%%%%%%%%%%%%	

\section{CONCLUSION}
In conclusion, we present a detailed study on the magnetic and electronic transport properties of the half-Heusler system MnPt$_{1-x}$Ir$_{x}$Sn. We find that the saturation magnetization systematically decreases while the anomalous Hall resistivity increases with increasing iridium concentration. The experimental Hall signal is analyzed using the scaling relation between anomalous Hall resistivity and longitudinal resistivity. We find that the scaling factor changes from close to 1 for the parent MnPtSn sample to quadratic for MnPt$_{0.5}$Ir$_{0.5}$Sn, signifying that Ir doping  enhances the intrinsic contribution by suppressing the extrinsic mechanism. Our experimental results are well supported by the theoretical study that showed that the Ir doping significantly enhances the spin-orbit coupling in the system.
The present study is an important contribution toward the basic understanding of different mechanisms of AHE, and thereby possesses a great importance in designing anomalous Hall sensor based spintronic devices.

%%%%%%%%%%%%%%%%%%%%%%%%%%%%%%%%%%%%%%%%%%%%%%%%%%%%%%%%%%%%%%%%%%%%%%%%%%%%%%%%%%%%%%%%%%%%%%%%%%%%%%%%%%%%%

\section{Acknowledgment}
 A.K.N. acknowledges support from the Department of Atomic Energy (DAE), the Department of Science and Technology (DST) Ramanujan Research Grant No. SB/S2/RJN-081/2016. R.R. acknowledges IIT Goa for her research fellowship and S.K. acknowledges DST INSPIRE for research funding.

%%%%%%%%%%%%%%%%%%%%%%%%%%%

\newpage
\section*{SUPPLEMENTARY INFORMATION}
\section{X-ray Diffraction}
\begin{figure}[h]
	\includegraphics[angle=0,width=8.5cm,clip]{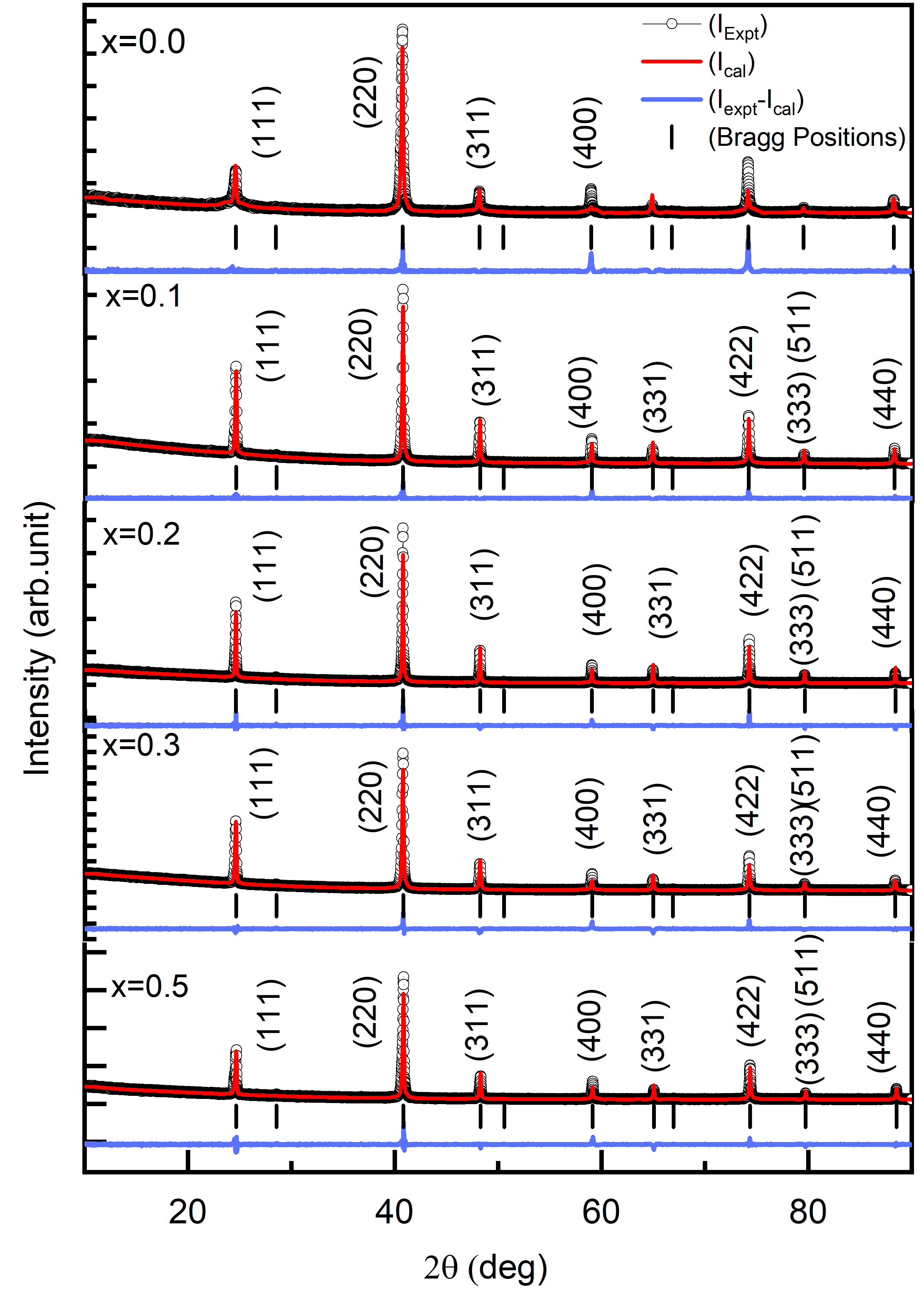}
	\caption{Room temperature powder XRD patterns and Rietveld refinements for MnPt$_{1-x}$Ir$_{x}$Sn (x=0, 0.1, 0.2, 0.3, 0.5
		\label{XRD}}
\end{figure}

X-ray diffraction (XRD) data are collected at room temperature with a Cu-K source on a Rigaku SmartLab x-ray diffractometer to check the phase purity of all the samples. The Rietveld refinement of the room temperature powder XRD patterns for MnPt$_{0.5}$Ir$_{0.5}$Sn are shown in FIG. \ref{XRD}. The Rietveld refinement for (x=0, 0.1, 0.2, 0.3) is performed using the Wyckoff positions as Sn at (0, 0, 0), Mn at (1/2, 1/2, 1/2), and Pt/Ir at (1/4, 1/4, 1/4). For x=0.5, same Wyckoff positions are used with 10\% atomic disorder between Mn/Ir sites (see section VII). The refinement confirms that all the samples are in single-phase and exhibit XRD patterns of cubic crystal structure of space group F$\bar{4}$3m.
\newpage
\section{Neutron Diffraction (ND)}
\begin{figure}[h]
	\includegraphics[angle=0,width=8cm,clip]{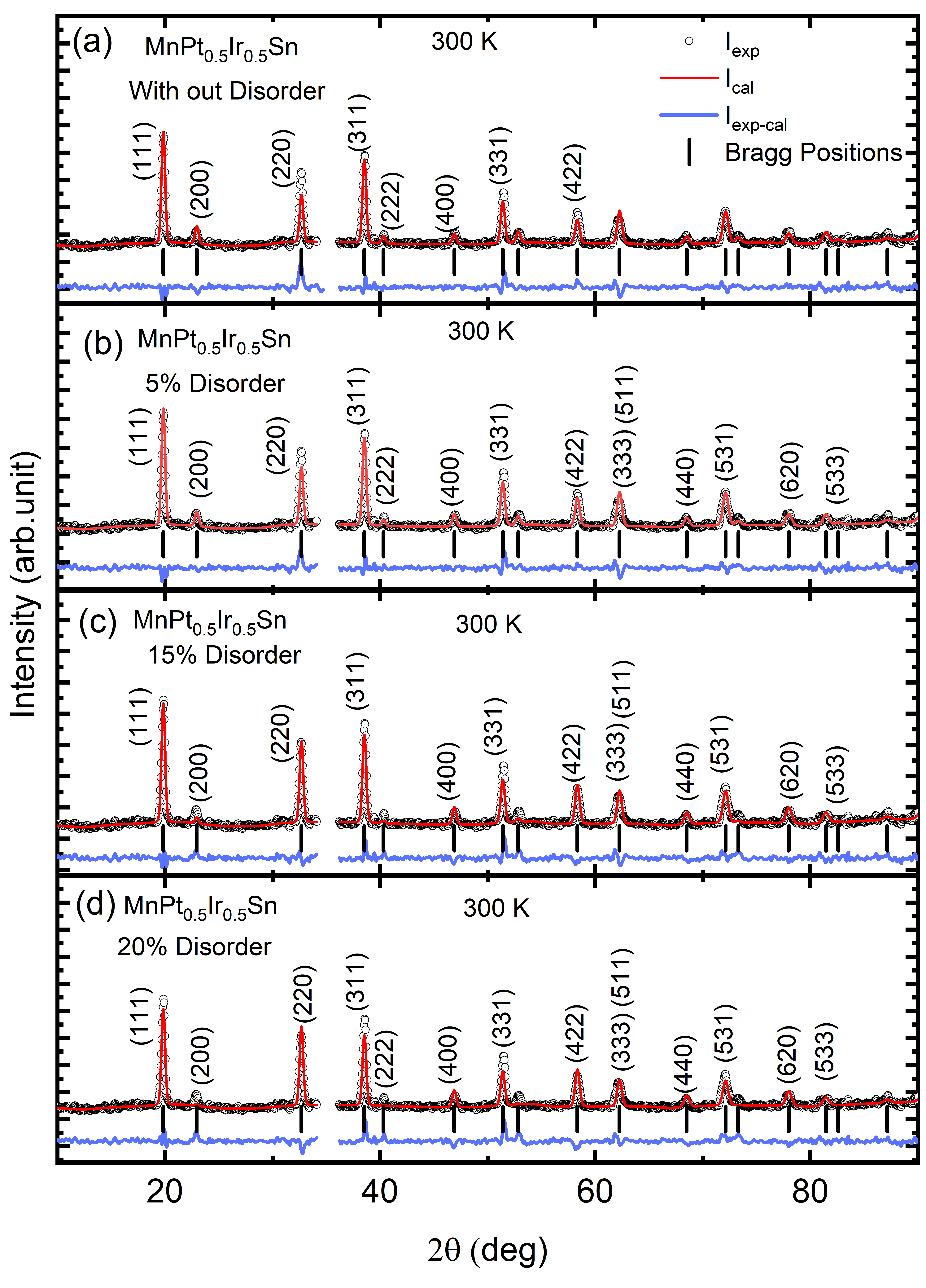}
	\caption{Rietveld refinement of the room-temperature ND pattern for MnPt$_{0.5}$Ir$_{0.5}$Sn with different atomic disorder between the Mn and Ir sites. The region between $34^o$ and $36^o$  is excluded due to contribution from the cryostat.
		\label{ND}}
\end{figure}

The room-temperature Rietveld refinement of the powder ND data for MnPt$_{0.5}$Ir$_{0.5}$Sn is carried out using the FullProf software with different percentages of atomic disorder between Mn and Ir. Figs. \ref{ND} (a)-\ref{ND} (d) shows the Rietveld refinement with 0\%, 5\%, 15\%, 20\% atomic disorders between the Mn and Ir atoms, respectively. The refined parameters with different percentages of atomic disorders are given in table II. We observe that 10\% atomic disorder between Mn/Ir results in a minimum value of $\chi^2$ and $R_{Bragg}$. The magnitude of $\chi^2$ and $R_{Bragg}$ increases as the disorder increase or decreases. Our findings indicate that the best fit for MnPt$_{0.5}$Ir$_{0.5}$Sn is obtained with 10\% atomic disorder between Mn/Ir sites.

\begin{table}[h]
		\caption{The percentage of disorder, site occupations, and other refined parameters for the room temperature powder neutron diffraction data of MnPt$_{0.5}$Ir$_{0.5}$Sn.}
		
		\label{tab:tableII}

	\begin{tabular}{|l|l|l|l|l|}
		\hline
		Sl. No        & \begin{tabular}[c]{@{}l@{}}Wyckoff \\ Position\end{tabular} & Site Occupations (\%)                                                            & $\chi^2$ & $R_{Bragg}$ \\ \hline
		0\% Disorder  & \begin{tabular}[c]{@{}l@{}}4b\\ \\ 4c\end{tabular}          & \begin{tabular}[c]{@{}l@{}}100 Mn + 0 Ir\\ \\ 50 Pt +50 Ir\end{tabular}          & 6.4  & 13.71  \\ \hline
		5\% Disorder  & \begin{tabular}[c]{@{}l@{}}4b\\ \\ 4c\end{tabular}          & \begin{tabular}[c]{@{}l@{}}95 Mn + 5 Ir\\ \\ 50 Pt +45 Ir + 5 Mn\end{tabular}    & 5.96 & 12.06  \\ \hline
		10\% Disorder & \begin{tabular}[c]{@{}l@{}}4b\\ \\ 4c\end{tabular}          & \begin{tabular}[c]{@{}l@{}}90 Mn + 10 Ir\\ \\ 50 Pt + 40 Ir + 10 Ir\end{tabular} & 5.96 & 11.62  \\ \hline
		15\% Disorder & \begin{tabular}[c]{@{}l@{}}4b\\ \\ 4c\end{tabular}          & \begin{tabular}[c]{@{}l@{}}85 Mn + 15 Ir\\ \\ 50 Pt +35 Ir + 15Ir\end{tabular}   & 6.42 & 14.38  \\ \hline
		20\% Disorder & \begin{tabular}[c]{@{}l@{}}4b\\ \\ 4c\end{tabular}          & \begin{tabular}[c]{@{}l@{}}80 Mn + 20 Ir\\ \\ 50 Pt +30 +Ir +20 Ir\end{tabular}  & 7.45 & 21.08  \\ \hline
	\end{tabular}
\end{table}

\newpage
\section{Temperature dependence of longitudinal resistivity}
\begin{figure}[h]
	\includegraphics[angle=0,width=8cm,clip]{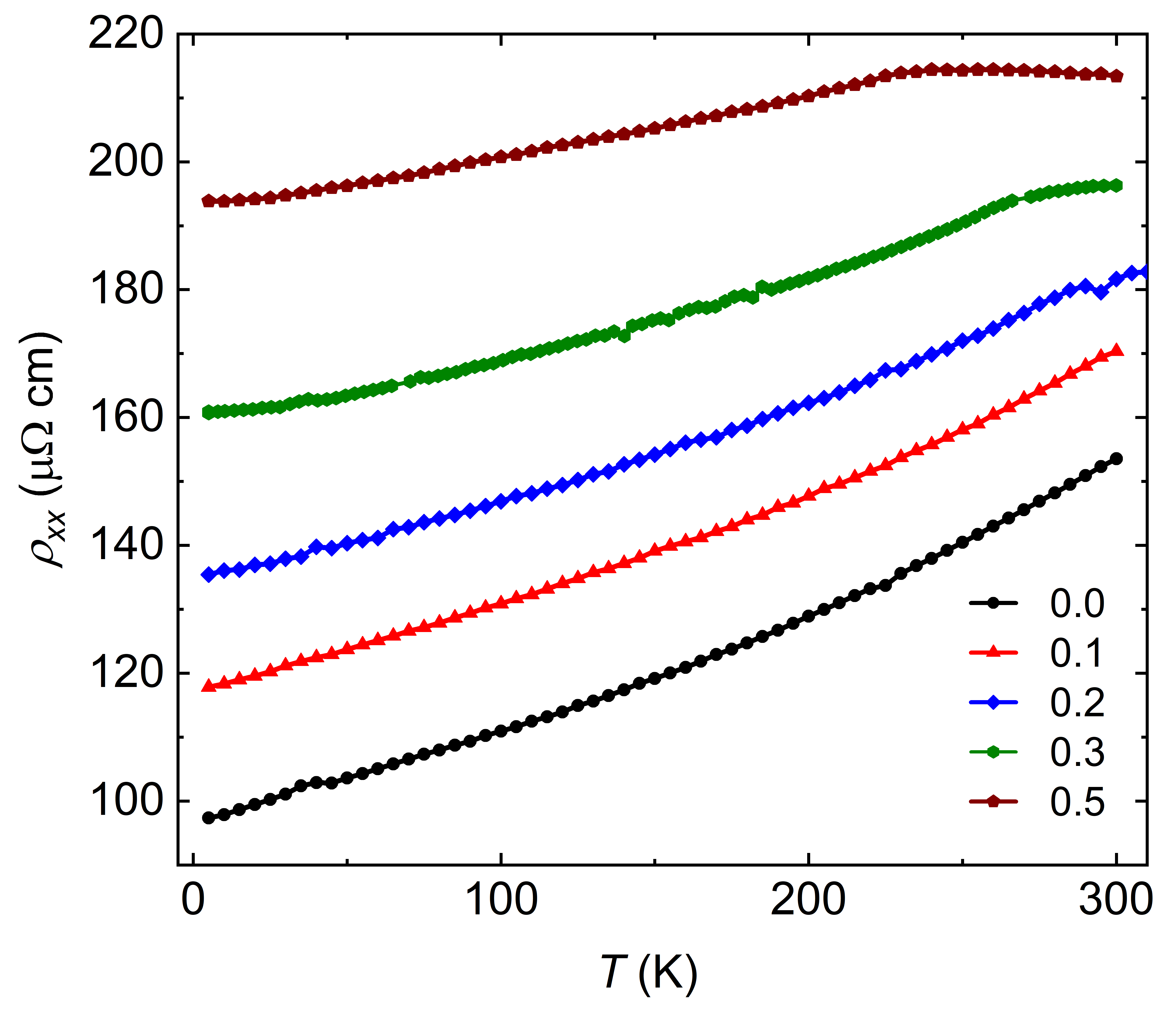}
	\caption{Temperature variation of longitudinal resistivity $\rho_{xx}$ (T) for the MnPt$_{1-x}$Ir$_x$Sn (with x= 0, 0.1, 0.2, 0.3, 0.5) samples.
		\label{RT}}
\end{figure}

Longitudinal resistivity $\rho_{xx}$ as a function of temperature (T) for all the samples is shown in Fig. \ref{RT}. Below the Curie temperature (Tc) resistivity decreases as the temperature decreases indicating that all the samples are in a metallic regime. At low temperature, $\rho_{xx}$ approaches the residual resistivity $\rho_{xx0}$
\section{Field dependent Hall resistivity}
\begin{figure}[h]
	\includegraphics[angle=0,width=8cm,clip]{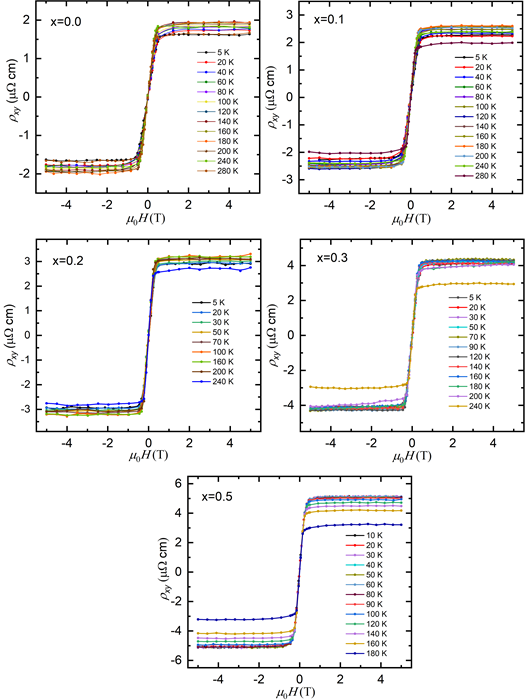}
	\caption{Field variation of Hall resistivity $\rho_{xy} (H)$ at different temperatures for the MnPt$_{1-x}$Ir$_x$Sn (with x= 0, 0.1, 0.2, 0.3, 0.5) samples.
		\label{S4}}
\end{figure}

The Hall resistivity measurements for all the samples are performed at different temperatures with field sweep $\pm$ 5T, as shown in Fig. \ref{S4}. For all the samples, the Hall resistivity increases with increasing temperature before it starts to decrease as the temperature approaches the Curie temperature.

\section{Density of states with the variation of spin-orbit coupling (SOC) strength}
\begin{figure}[h]
	\includegraphics[angle=0,width=8.5cm,clip]{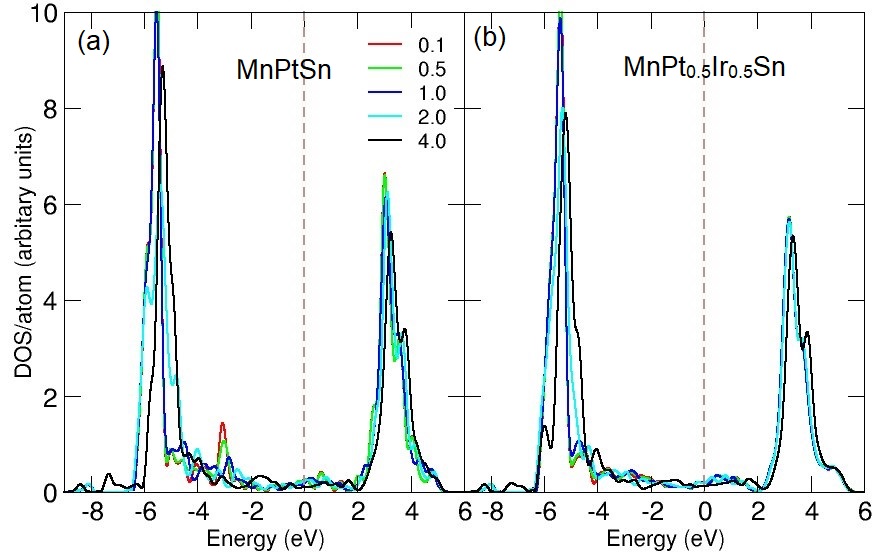}
	\caption{Projected density of states of Mn-d with the variation of spin orbit coupling (SOC) strength for (a) MnPtSn and (b) MnPt$_{0.5}$Ir$_{0.5}$Sn
		\label{S5}}
\end{figure}

The density of states for Mn-d states of MnPtSn and MnPt$_{0.5}$Ir$_{0.5}$Sn with the variation of spin orbit-coupling (SOC) strength are calculated and shown in Figs. \ref{S5} (a)-\ref{S5} (b), respectively. The calculated density of states with different SOC strengths does not show any significant changes for both the samples.
\section{Magnetic Exchange Interactions}
\begin{figure}[tb!]
	\includegraphics[angle=0,width=8.5cm,clip]{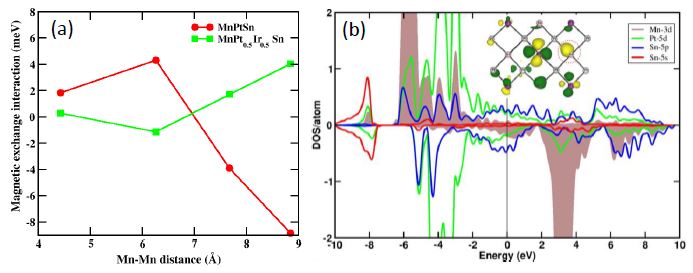}
	\caption{(a) Magnetic exchange interactions between different Mn sites as a function of the Mn-Mn distances for the MnPtSn and MnPt0.5Ir0.5Sn. Positive and negative signs represent ferromagnetic and antiferromagnetic type of the interactions, respectively. (b) Density of states (DOS) for MnPtSn. The Mn-3d, Pt-5d, Sn-5p, Sn-5s are represented by gray, green, blue and red respectively. Inset shows the Wannier function plot.
		\label{RKKY}}
\end{figure}
With the Ir doping, substantial changes are observed in the magnetic properties of the MnPtSn half Heusler compound. In the experiments, it is observed that Mn spins are ferromagnetically ordered in the case of MnPtSn, however, there is some sort of antiferromagnetic correlation in the Ir doped cases. Through the electronic structure total energy method, we calculate the exchange interactions between different Mn sites as a function of the Mn-Mn distances as shown in Fig. \ref{RKKY} (a). We find that the magnetic exchange interactions of the MnPt$_{0.5}$Ir$_{0.5}$Sn are weaker than the MnPtSn, which is consistent with the smaller value of the TC for the Ir doped samples compared to that of the parent MnPtSn. We also find that the type of the exchange interactions are oscillatory in nature which ensures the conduction electron-driven RKKY type of ferromagnetic exchange interactions are dominating in these materials. In the case of MnPtSn, the Mn-Mn distance is large; thus the Mn-3d state does not overlap directly, and the direct exchange interactions can be neglected. On the contrary, the RKKY type of exchange interaction between two magnetic Mn ions is also mediated by the itinerant conduction electrons. The itinerant electrons close to the Fermi energy and having substantial hybridization with the Mn-d states will act as a mediator in the RKKY interactions. For the MnPtSn, the density of states (Fig Fig. \ref{RKKY} (b)) clearly shows that Sn-sp mixed states are taking part in the conduction electron near the Fermi energy and acting as a mediator for the RKKY type exchange interactions between two Mn-d ions. We also found a small presence of Pt/Ir-5d states in the Mn-Mn exchange interactions. The below DOS and Wannier function (inset figure Fig. \ref{RKKY} (b)) plot clearly shows that the finite tail (marked by red dotted circle in the inset) situated at the Sn site with a shape of sp hybridized state as a mediator between Mn-Mn interactions. We also found a very tiny tail at the Pt site, which also has little contribution as a mediator.

\section{Calculation of Magnetic transition temperature}
The magnetic transition temperatures in a mean-field approach for both the systems are calculated using the equation
\begin{equation}
	T_C=\frac{2}{3} S(S+1) \sum_i J_iZ_i
\end{equation}
The experimental ratio of $T_C$ is $\frac{T_{C}^{MPS}}{T_{C}^{MPIS}}$ =1.44 (MPS represents for MnPtSn and MPIS represents for MnPt0.5Ir0.5Sn). 
The calculated ratio of $T_C$ is $\frac{T_{C}^{MPS}}{T_{C}^{MPIS}}$ =1.52.
\section{Side Jump Contribution}
\begin{figure}[h]
	\includegraphics[angle=0,width=7cm,clip]{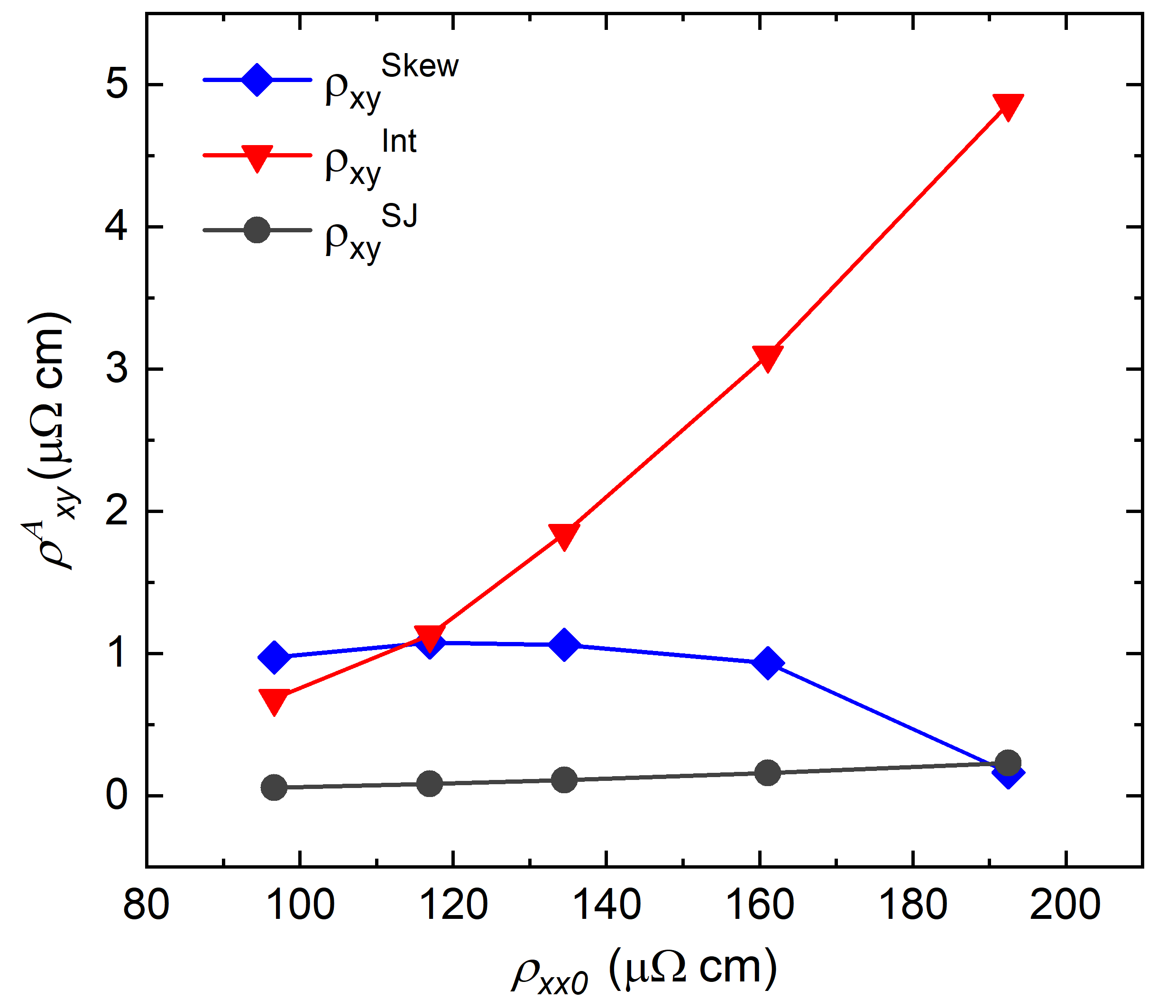}
	\caption{Skew scattering $\rho_{xy}^{Skew}$, intrinsic  $\rho_{xy}^{Int}$ and side-jump  $\rho_{xy}^{SJ}$ contributions with respect to residual resistivity  $\rho_{xx0}$ for MnPt$1-x$Ir$_x$Sn (for x = 0, 0.1, 0.2, 0.3, 0.5).
		\label{SJ}}
\end{figure}
According to the existing theoretical model, the extrinsic side-jump contribution is in the order of $\sigma_{xy}^{SJ}\approx\frac{e^2}{ha} (\frac{\epsilon_{SO}}{E_F})$ , where $\epsilon_{SO}$ is the spin-orbit interaction energy, $E_F$ is the Fermi energy, a is the lattice constant. For metallic ferromagnet $\frac{\epsilon_{SO}}{E_F} \approx 0.01$. With the help of this expression, one can have the order of side jump contribution and compares it with the other mechanism. We calculated the approximate side jump contribution for our system using the above expression and taking the experimental lattice parameter. We plotted the different contributions with residual resistivity $\rho_{xx0}$ . It is clearly seen that the SJ contribution is minimal compared to other dominant contributions.

\end{document}